\documentclass[reprint,superscriptaddress,floatfix,amsmath,amssymb]{revtex4-1}
\usepackage{graphicx}
\begin{document}
\newcommand{\ket}[1]
{|#1 \rangle}
\newcommand{\bra}[1]
{\langle #1 |}
\newcommand{\singlet}[0]
{{}^{1}S_{0}}
\newcommand{\triplet}[0]
{{}^{3}P_{2}}
\newcommand{\gup}[0]
{g_{\uparrow}}
\newcommand{\gdown}[0]
{g_{\downarrow}}
\newcommand{\gnd}[0]
{g}
\newcommand{\gupdown}[0]
{g_{\uparrow,\downarrow}}
\newcommand{\exc}[0]
{e}

\title{Spatial Adiabatic Passage of Massive Quantum Particles}
\author{Shintaro Taie}
\altaffiliation{Electronic address: taie@scphys.kyoto-u.ac.jp}
\affiliation{Department of Physics, Graduate School of Science, Kyoto University, Japan 606-8502}
\author{Tomohiro Ichinose}
\affiliation{Department of Physics, Graduate School of Science, Kyoto University, Japan 606-8502}
\author{Hideki Ozawa}
\affiliation{Department of Physics, Graduate School of Science, Kyoto University, Japan 606-8502}
\author{Yoshiro Takahashi}
\affiliation{Department of Physics, Graduate School of Science, Kyoto University, Japan 606-8502}
\date{\today}

\begin{abstract}
By adiabatically manipulating tunneling amplitudes of cold atoms in a periodic potential with a multiple sublattice structure,
we are able to coherently transfer atoms from a sublattice to another without populating the intermediate sublattice,
which can be regarded as a spatial analogue of stimulated Raman adiabatic passage.
A key is the existence of dark eigenstates forming a flat band in a Lieb-type optical lattice.
We also successfully observe a matter-wave analogue of Autler-Townes doublet using the same setup.
This work shed light on a novel kind of coherent control of cold atoms in optical potentials.
\end{abstract}
\maketitle
Interference of probability amplitudes is one of the most significant properties of quantum mechanics.
In the seminal work of an electron double-slit experiment \cite{Tonomura1989}, building up of the interference pattern of 
electron wave function beautifully demonstrated nature of wave-particle duality.
Quantum interference has been intensively utilized especially in the field of precision measurement
such as superconducting magnetometer \cite{Jaklevic1964} and atom interferometry \cite{Muller2010},
and also lies at the heart of quantum information science.

\begin{figure}[bt]
	\includegraphics[width=85mm]{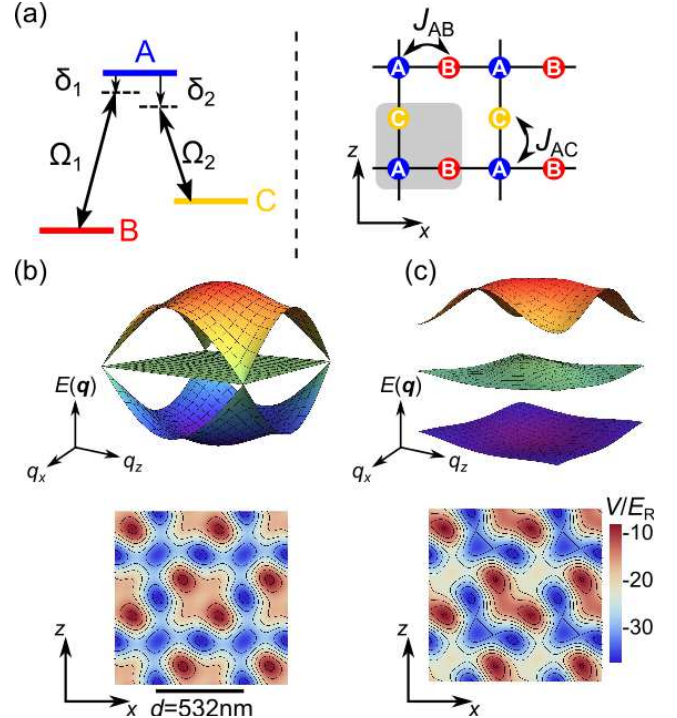}
	\caption{
		(a) A $\Lambda$-type three level system and a Lieb lattce. Gray shaded region indicates one of unit cells.
		(b) Energy band structure of the Lieb lattice with no distortion, calculated in the tight-binding limit.
		The corresponding potential landscape in our optical lattice setup is also shown below. 
		(c) Energy band and potential landscape with distortion. If the phase of the diagonal lattice is shifted,
		band gaps become large and the Dirac cone disappears.
		Shown is the case of $\psi=(1/2+0.11)\pi$ and $s=[(8, 8), (8, 8), 14]$, which is used in Fig. \ref{fig_sap} (a).
	}
	\label{fig_schem}
\end{figure}

A three-level system is a minimal example in which quantum interference takes place.
Most commonly it is considered in a context of laser coupled atomic levels, and the Hamiltonian
for a $\Lambda$-type system (Fig. \ref{fig_schem} (a)) in a rotating frame is written in the form
\begin{equation}
H =
	\left( 
		\begin{array}{ccc}
			0				&	\Omega_1	&	\Omega_2\\
			\Omega_1	&	\delta_1	&	0\\
			\Omega_2	&	0	&	\delta_2
		\end{array} 
	\right), \label{eq_Hamiltonian1}
\end{equation}
where $\Omega_{1}$ ($\Omega_{2}$) denotes a laser-induced Rabi frequency which couples basis states
$\ket{A}$ with $\ket{B}$ ($\ket{A}$ with $\ket{C}$), and $\delta_1$ ($\delta_2$) is the detuning of
the corresponding laser 1 (2).
A dark state $\cos \theta \ket{B}-\sin \theta \ket{C}$ ($\tan \theta = \Omega_1/\Omega_2$) arises as one of
the eigenstate of the Hamiltonian given by Eq. (\ref{eq_Hamiltonian1}) if the Raman resonant condition
$\delta_1 = \delta_2$ is satisfied.
By applying two laser pulses in so-called ``counter-intuitive'' order so that $\theta$ changes form $0$ to $\pi/2$,
the dark states smoothly evolve from $\ket{B}$ into $\ket{C}$. This process is well known as Stimulated Raman Adiabatic Passage
(STIRAP) \cite{Kuklinski1989,Gaubatz1990,Bergmann1998}, and has been an important technique for robust population transfer
between two atomic/molecular states.

Natural interests arise for a special case that the states of interest represent a matter wave of a quantum particle and
the initial and the final states are spatially well isolated.
Such processes, named spatial adiabatic passage (SAP), offer paradoxical ``transport without transit'' \cite{Rab2008,Benseny2012}
where matter waves are transported without populating the intermediate spatial region. 
Since the concept of SAP was introduced in the context of quantum dots \cite{Renzoni2001,Greentree2004}
and cold atoms \cite{Eckert2004}, it has continuously attracted theoretical interests and various possibilities
of its application have been discussed \cite{Menchon2016}.
Generalized to two-dimensions, SAP technique also enables preparation of states with angular momentum \cite{McEndoo2010,Menchon2014a}. 
It is also interesting to consider the case of interacting many-body systems such as Bose-Einstein condensates \cite{Graefe2006,Rab2008,Rab2012}.
Highly controllable and flexible systems of cold atoms are suitable to realize SAP and the above-mentioned applications.

A key for the realization of SAP is the existence of dark eigenstates.
Interestingly, dark states also arise in eigenstates for a particle moving in a special kind of lattice structures.
In this paper, we report on the successful realization of SAP by adiabatic control of matter-wave tunneling in an optical lattice.
Our lattice consists of three sublattices ($A$, $B$, and $C$ shown in Fig. \ref{fig_schem} (a)), forming a Lieb lattice.
It is convenient to take plane waves on each sublattice $\ket{\vec{q}, A}$, $\ket{\vec{q}, B}$ and $\ket{\vec{q}, C}$
with a momentum $\vec{q}$ as a basis set.
The existence of nearest-neighbor tunneling $J_{AB}$ and $J_{AC}$ induces coupling among these basis states
given by ${\cal T}_{AB}(\vec{q}) = -2J_{AB} \cos(q_xd/2)$ and ${\cal T}_{BC} = -2J_{BC} \cos(q_zd/2)$.
The resulting Hamiltonian can be written as a $3\times 3$ matrix form
\begin{equation}
H = \sum_{\vec{q}} H_{\vec{q}},\ \ \ 
H_{\vec{q}} =
\left( 
\begin{array}{ccc}
E_A	&	{\cal T}_{AB}(\vec{q})	&	{\cal T}_{AC}(\vec{q})\\
{\cal T}_{AB}(\vec{q})	&	E_B	&	0\\
{\cal T}_{AC}(\vec{q})	&	0	&	E_C
\end{array} 
\right). \label{eq_Hamiltonian2}
\end{equation}
Now analogy to a $\Lambda$-type system (\ref{eq_Hamiltonian1}) is obvious: 
momentum-dependent couplings play a role of Rabi couplings in a three-level system and
detunings can be mimicked by energy offsets $E_A$, $E_B$, and $E_C$ of each sublattice.
All these parameters can be controlled by changing the lattice depth along each direction,
which enables us to realize a coherent scheme to transport atoms among these sublattices.
Note that classical waves traveling through coupled waveguides is known to obey coupled mode equations \cite{Hardy1985}
which are mathematically equivalent to Eq. (\ref{eq_Hamiltonian1}) and SAP correspondence was demonstrated \cite{Menchon2012}.
However, the original concept, SAP with matter waves of a quantum particle has not been realized.
\section{Expermental Realization}
Our experimental setup is similar to that described in Ref. \cite{Taie2015,Ozawa2017}. In brief, an optical Lieb lattice is created by
the combination of a 2D staggered superlattice with $d/2=266$~nm spacing and a diagonal lattice with
$\sqrt{2} \times 266$~nm periodicity. The resulting potential $V(\vec{x})$ is given by
$V(\vec{x})/E_R = - s_{\rm short}^{(x)} \cos^2 (2\pi x/d) - s_{\rm short}^{(z)} \cos^2 (2\pi z/d) - s_{\rm long}^{(x)} \cos^2 (\pi x/d)-s_{\rm long}^{(z)} \cos^2 (\pi z/d) - s_{\rm diag} \cos^2 [2\pi (x-z)/d + \psi]$,
where $E_R = \hbar^2/2m (\pi/d)^2$ is the recoil energy for the long lattice.
Below we specify a lattice potential by $s$-parameters $s = [(s_{\rm short}^{(x)}, s_{\rm short}^{(z)}), (s_{\rm long}^{(x)},s_{\rm long}^{(z)}), s_{\rm diag}]$ and the relative phase of the diagonal lattice $\psi$.
Basically, tunneling amplitudes are determined by the depths of short lattices ($s_{\rm short}^{(x)}$, $s_{\rm short}^{(z)}$)
and the other lattice depths are responsible for the energy offsets $E_A$, $E_B$ and $E_C$.
In our experiment, we use fermionic $^{171}$Yb with a small scattering length $-0.14$~nm to avoid
interaction effects. The use of fermions introduces a complexity arising from the finite momentum spread
due to the Pauli principle. In the absence of interactions and harmonic confinement, the dynamics conserves
quasimomentum and states initially having well-defined quasimomentum $\vec{q}$ evolve within a subspace
spanned by three plane waves $\ket{\vec{q}, A}$, $\ket{\vec{q}, B}$ and $\ket{\vec{q}, C}$.

Adiabaticity of a process associated with a certain momentum is governed by the band gaps among the corresponding eigenstates.
This implies that, for a Lieb lattice, adiabaticity cannot be maintained around the corner of the Brillouin zone
where a Dirac cone exists (Fig. \ref{fig_schem} (b)). To overcome this problem, we slightly deform the lattice structure
by shifting the phase from an isotropic condition $\psi = \pi/2$. The effectiveness of this scheme can be understood
from the potential landscape shown in Fig. \ref{fig_schem} (c).
The deformation reduces the inter-unit-cell tunneling, therefore each cell becomes more like an isolated triple well.
As a result, the momentum dependence of the dispersion curves is reduced and the Dirac cone is eliminated.
Mathematically, this modifies the coupling term as
${\cal T}_{AC} \rightarrow e^{i q_zd/2} (J_{AC}+\delta J) + e^{-i q_zd/2} (J_{AC} - \delta J)$
(similar change applies for ${\cal T}_{AB}$),
where $\delta J$ denotes the imbalance between inter- and intra-unit-cell tunneling.
For $\delta J =0$, ${\cal T}_{AC}$ along the Brillouin zone boundary ($q_zd =\pi$) vanishes throughout the process.
Introduction of $\delta J \neq 0$ can also suppress the breakdown of transport along this line. 

\section{Spatial Adiabatic Passage}
\begin{figure}[bt]
	\includegraphics[width=85mm]{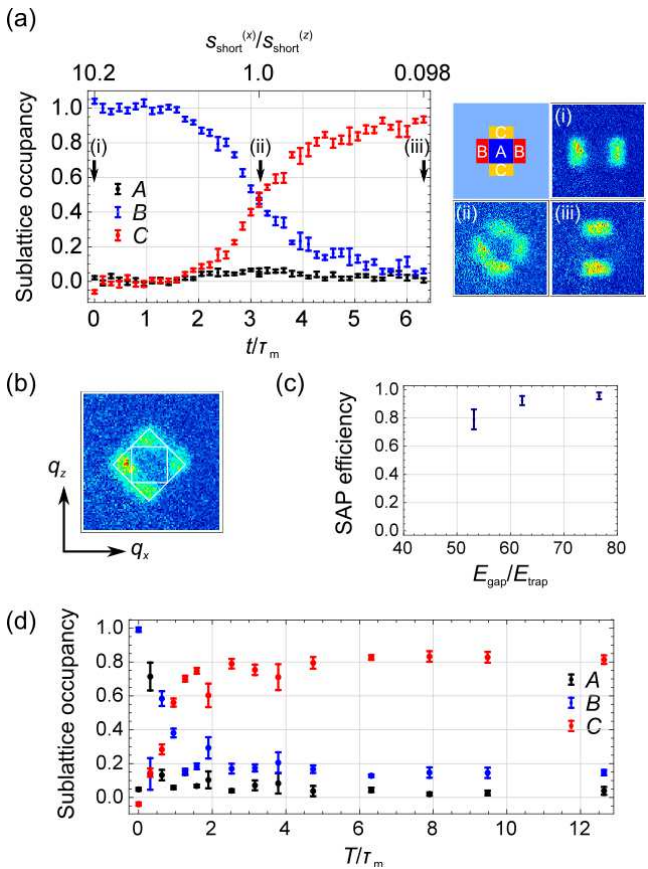}
	\caption{
		(a) Time evolution of sublattice occupancies $N_A$, $N_B$ and $N_C$ during a SAP process.
		Time is rescaled by $\tau_{\rm m}=0.63$ ms.
		Samples of absorption images used taken in the experiment are also shown in the right hand side.
		Labels (i)-(iii) represent the correspondence between images and data in the graph.
		(b) Band-mapping image at the half point (ii) of the SAP process (a). Boundaries for the first and second Brillouin zone
		are indicated by white lines.
		(c) Dependence of SAP efficiency on the ratio $E_{\rm gap}/E_{\rm trap}$.
		(d) Adiabaticity of the SAP process. Sublattice occupancies after SAP are shown as a function of total sweep time $T$.
	}
	\label{fig_sap}
\end{figure}
A matter-wave analogue of a STIRAP in the Lieb lattice is to transport atoms between two sublattices ($B$ and $C$)
by a counter-intuitive temporal change of tunneling amplitudes. Throughout this process, the intermediate sublattice
$A$ is not populated because the state adiabatically follows a dark state
$\cos \theta_{\vec{q}} \ket{\vec{q}, B}- \sin \theta_{\vec{q}} \ket{\vec{q}, C}$, with
$\tan \theta_{\vec{q}} = {\cal T}_{AB}(\vec{q})/{\cal T}_{AC}(\vec{q})$.
First we load a sample of $1.2 \times 10^4$ atoms at a temperature as low as $0.3$ of the Fermi temperature
into the optical Lieb lattice.
In the loading stage, the potential on a $B$-sublattice is made much deeper than those of the others
($s = [(27.7, 0), (0, 16), 14]$) to ensure that the initial state is only populated by $B$.
After that, we quickly change the lattice depths to $[(38.9, 3.8), (8, 8), 14]$ in $10$ $\mu$s.
This is a starting point of STIRAP process, where the tunneling $J_{AB}$ is much smaller than $J_{AC}$.
To achieve a high tunneling rate, overall lattice depths are set relatively shallow, leading to an unwanted direct
tunneling $J_{BC}$. We suppress $J_{BC}$ by increasing the diagonal lattice depth beyond the equal-offset condition,
i.e., $E_A > E_B=E_C$. As long as $E_B=E_C$ is maintained,
the dark state persists and STIRAP can be accomplished. We adiabatically sweep the lattice depths toward another limiting
configuration $[(3.8, 38.9), (8, 8), 14]$, passing through the intermediate point $[(8, 8), (8, 8), 14]$ corresponding to
the potential shown in Fig. \ref{fig_schem} (c).
The time evolution during this process is monitored by mapping sublattice populations onto band populations 
followed by a standard band mapping technique \cite{Taie2015,Ozawa2017}. 
The obtained time-of-flight images suffers from the blurring of the Brillouin zone boundaries due to unavoidable
non-addiabaticity of the band mapping procedure and a harmonic confinement of the system.
For a qualitative analysis of sublattice populations, we take a set of basis images
in which all atoms resides on a specific sublattice and determine sublattice occupancies by projecting images onto each basis.
Figure \ref{fig_sap} (a) shows the time evolution of sublattice occupancies $N_A$, $N_B$ and $N_C$ during the SAP process.
The time scale is renormalized by $\tau_{\rm m}=2\pi \hbar/E_{\rm gap}$ where $E_{\rm gap}$ is the energy gap averaged over
entire Brillouin zone at the half point of the process.
Important features specific to SAP are well reproduced: initial population on the $B$-sublattice, $N_B$, is smoothly transfered
into $N_C$, but $N_A$ does not show increase throughout the process. From the final population we evaluate the efficiency
of the process to be $E = [N_C(T)-N_C(0)]/N_B(0) = 0.95(2)$.
During the SAP process the state is certainly following the dark state as one can see in the direct band-mapping image
shown in Fig. \ref{fig_sap} (b). Here, the concentration of the atomic distribution inside the second Brillouin zone
indicates the state is kept in the second energy band which consists of the dark states.
Usually, occupation of a certain energy band is accomplished by filling up all lower bands.
The above SAP process provides an efficient way to prepare a non-equilibrium many-body state in which
all fermions reside on the flat band of the Lieb lattice and the other bands are empty.

In our system, the SAP efficiency is limited by the existence of a trap-induced harmonic confinement.
This causes inhomogeneity of energy offsets, especially around the edge of an atomic cloud.
We change the ratio of $E_{\rm gap}$ to $E_{\rm trap}=m \omega^2 d^2/2$ with $\omega$ denoting
the mean trap frequency on $x-z$ plane, and monitor the change of SAP efficiency.
As seen in Fig. \ref{fig_sap} (c), the trap reduces the SAP efficiency.

We examine adiabaticity of the STIRAP process by changing a total sweep time $T$(Fig. \ref{fig_sap} (d)).
As naively expected, for $T/\tau_{\rm m} \lesssim 1$, adiabaticity is broken and significant populations in $A$ and $B$
are observed. 
Once the adiabatic condition is fulfilled for large $T/\tau_{\rm m}$, the final state is kept almost constant,
with a large population in $C$ regardless of $T$. This behavior is characteristic to a robust adiabatic process,
in contrast to Rabi oscillations driven by a direct tunneling coupling.

\section{Bright State Transport}
\begin{figure}[bt]
	\includegraphics[width=85mm]{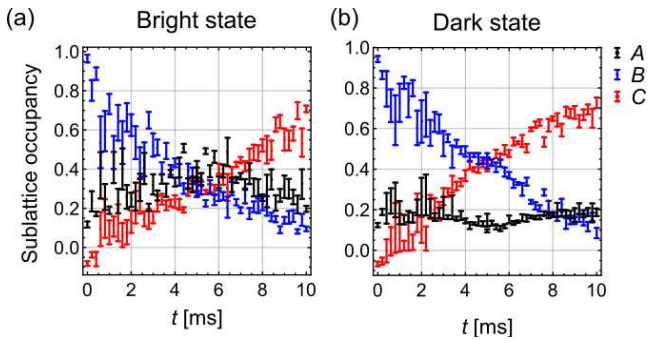}
	\caption{
		(a) Transport from the $B$- to $C$-sublattice through the bright state.
		(b) Transport through the dark state. After loading to the $B$-sulattice, we exactly reverse
		the potential sweep applied in (a). 
	}
	\label{fig_bright}
\end{figure}

In atomic three-level systems on the one hand, an intermediate $\ket{A}$ state generally suffers from significant loss
due to spontaneous emission.
One of the advantages for STIRAP is that the loss through $\ket{A}$ state can be avoided by keeping the dark state.
On the other hand, SAP with cold atoms does not suffer from any losses through intermediate states, which enables
us to perform efficient transport through ``bright'' states where $\ket{A}$ is significantly populated.
We design an ``intuitive'' potential sweep for this bright state transport for the optical Lieb lattice.
The sweep $[(25,30),(7.6,8.4),14]  \rightarrow [(8,8),(8,8),14] \rightarrow [(30,25),(8.4,7.6),14]$
involves not only $s_{\rm short}^{(x,z)}$ (tunneling) but also $s_{\rm long}^{(x,z)}$ (detuning)
to improve the transport efficiency.
Figure \ref{fig_bright} (a) shows the resulting time evolution during this sweep.
Due to the requirement of minimizing unwanted excitations at the beginning of the sweep,
initial localization on $B$ is not perfect in this experiment. However, the nature of the bright state transport is clearly
visible as the increase of $N_A$ around the half point.
As a reference, we also perform a ``counter-intuitive'' scheme under the same condition.
In Fig. \ref{fig_bright} (b) we start with a sample localized on $B$, but apply the potential sweep in a time-reversal way
compared with that of \ref{fig_bright} (a) to transport atoms through the dark state.
This scheme is essentially equivalent to the SAP process demonstrated in the previous section,
except that the state does not always remain dark because $E_B-E_C$ changes during sweep.
Similar performance of transport efficiency is obtained, but the
behavior around the half point of the process is qualitatively different. At this point, the state becomes exactly dark
and the $N_A$ certainly shows its minimum, in contrast to the bright state transport.
The feasibility of accessing both bright and dark state transfer manifests the flexibility of our system in quantum state engineering.

\section{Autler-Townes doublet}
Electromagnetically induced transparency (EIT) \cite{Boller1991} is also an important process in a three-level system.
In an EIT experiment, there is strong optical coupling between $A$ and $C$ which causes splitting of the $B \rightarrow A$
transition by the Rabi frequency, known as Autler-Townes doublet \cite{Autler1955}. As a result, the state $B$ becomes
transparent for laser light driving the $B \rightarrow A$ transition at a frequency region between the doublet.

\begin{figure}[bt]
	\includegraphics[width=85mm]{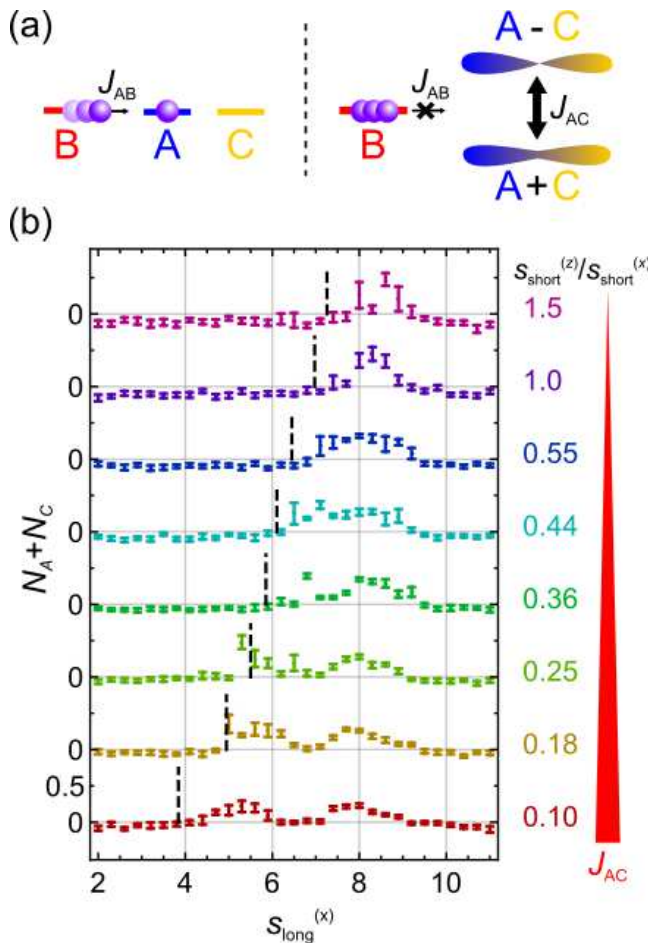}
	\caption{
		(a) Schematic of the experiment.
		(left) In the absence of $J_{AC}$, the weak tunneling coupling $J_{AB}$ results in a fraction of
		atoms in $A$.
		(right) In the presence of strong $J_{AC}$, tunneling only occurs when the energy resonates to
		bonding $A+C$ or anti-bonding $A-C$ orbitals, split by coupling $J_{AC}$.
		(b) Tunneling spectrum after a hold time of $1.8$ ms, at the lattice depths of
		$[(40,s_{\rm short}^{(z)}),(s_{\rm long}^{(x)},8),9.5]$.
		Fraction of atoms which tunnel from the $B$-sublattice is shown.
		Vertical dashed lines show the rough estimation of the points where $E_A=E_B$ is satisfied,
		based on a harmonic approximation of the lattice potential.
	}
	\label{fig_eit}
\end{figure}

To investigate a matter wave analogue of EIT physics, we carry out a measurement similar to a pump-probe experiment,
as depicted in Fig. \ref{fig_eit} (a).
As before, we first prepare an initial state localized on $B$. Then we allow weak tunneling coupling
$J_{AB}$ and after a fixed time, a fraction of atoms that tunneled into $A$ or $C$ is measured.
Figure \ref{fig_eit} (b) shows a set of tunneling spectra. In each spectrum we scan the ``detuning'' $s_{\rm long}^{(x)}$
which determines the energy difference $E_B - E_A$ ($=E_B -E_C$). As we increase the coupling $J_{AC}$
(decrease $s_{\rm short}^{(z)}$), the spectrum drastically changes: for negligible $J_{AC}$, we can observe
a single peak corresponding to $B \rightarrow A$, whereas a clear doublet structure appears for $J_{AC} \gg J_{AB}$.
The double peaks originate from tunnelings to $A+C$ and $A-C$ orbitals which are separated by the 
amplitude of tunneling coupling $J_{AC}$.
The overall shift of the spectrum is caused by the change of the short lattice depth $s_{\rm short}^{(z)}$.
While the short lattice creates the same potential curve for all sublattices, its effect on $E_A$, $E_B$ and $E_C$
slightly differs depending on the configuration of other lattices. We estimate the shift of zero-point energy by a harmonic
approximation. Whereas it fails to predict the position of resonance center, the tendency of the shift is well captured.
The width of the observed resonance peaks is broadened by band dispersion and spatial inhomogeneity.
In a typical EIT spectrum, a sharp dip can be observed even when the doublet splitting is smaller than
the natural linewidth. This implies occurrence of coherent population trapping (CPT) \cite{Arimondo1996} of a dark eigenstate.
In the case of our system with no loss mechanism, CPT does not occur and hence a sharp EIT dip does not appear.
However, the observed behavior exactly corresponds to a pump-probe
detection of Autler-Townes doublet which is commonly observed in atom-field systems.
As a future direction, it is interesting to introduce site-dependent loss (by e.g., high-resolution laser spectroscopy)
and study CPT and related phenomena.
		
\section{Conclusion}
In conclusion, we have succeeded in demonstrating coherent tunneling processes of cold atoms in an optical Lieb lattice.
The three-sublattice structure of the Lieb lattice has remarkable analogy to $\Lambda$-type three level systems in quantum optics.
By using this analogy and dynamical controllability of tunneling amplitudes, spatial addiabatic passage between two sublattice
eigenstates was performed. We also observed an matter-wave analogue of Autler-Townes doublet in a tunnleing process
from a sublattice into a strongly coupled pair of sublattices. The demonstrated techniques are useful to prepare
exotic many body states in optical lattices. For example, at the half point of the SAP process in the Lieb lattice,
all atoms are located on the flat band.
This might be a general scheme applicable to other lattice with flat bands.
Involving higher lattice orbitals is also interesting in connection with generation of angular momentum studied in Ref. \cite{Menchon2014a}.
In addition, recent advances in fine potential engineering \cite{Liang2010} combined with site-resolved imaging of lattice gases \cite{Bakr2009}
will greatly enlarge the application of SAP in cold atomic systems. 

\begin{acknowledgments}
We thank, T. Busch, J. Mompart, A. Greentree and V. Nesterenko for stimulating discussions, and H. Shiotsu for experimental help.
This work was supported by the Grant-in-Aid for Scientific Research of JSPS
(No. JP25220711, No. JP26247064 No. JP16H00990, No.JP16H01053), the Impulsing Paradigm Change through Disruptive Technologies
(ImPACT) program, and JST CREST(No. JPMJCR1673). H.O. acknowledges support from JSPS Research Fellowships.
\end{acknowledgments}
\section{Supplementary Information}
	\subsection{Sample Preparation}
		A quantum degenerate gas of $^{171}$Yb is produced by sympathetic evaporative cooling
		with fermionic $^{173}$Yb, in a crossed dipole trap operating at $532$~nm. \cite{Taie2010}.
		After evaporation, remaining $^{173}$Yb atoms are cleaned up by illuminating resonant laser light
		on the ${}^1S_{0} \leftrightarrow {}^3P_1 (F=7/2)$ transition. 
		All the data presented in the main text are taken with spin unpolarized samples.
	\subsection{Optical Lieb Lattice}
		Our optical Lieb lattice consists of a short square lattice operating at a wavelength of $532$~nm,
		a long square lattice at $1064$~nm and a diagonal lattice at $532$~nm.
		The lattice beams for two square lattices are retroreflected by common mirrors.
		Therefore the relative phases between these lattices are determined by the relative frequencies between
		the short and long laser beams, which is actively stabilized to	realize the staggered lattice configuration.
		The relative phase $\psi$ of the diagonal lattice is stabilized by a Michelson interferometer
		and the uncertainty is estimated to be $0.02\pi$ during a typical measurement time \cite{Taie2015}.
	\subsection{Sublattice occupation measurent}
		To project sublattice occupations onto band occupations, the lattice potential is suddenly changed into
		$s=[(8, 8), (8, 8), 0]$. Without the diagonal lattice, the $x$ and $z$ directions are decoupled and
		atoms in $B$ ($C$) sublattice is mapped onto the 1st excited band of the $x$ ($z$) direction.
		Followed by the band mapping technique, atoms distribute over the ($1+1$) dimensional Brillouin zone
		with the one-to-one correspondence to sublattice occupancies ($N_A$, $N_B$ and $N_C$).

		To analyze data, we fit an empirical model function
		$F(q_x, q_z) = N_A f_A (q_x, q_z) + N_B f_B (q_x, q_z) + N_C f_C (q_x, q_z) + b$
		with taking $N_A$, $N_B$, $N_C$ and the background level $b$ as free parameters.
		Here, the basis functions $f_A$, $f_B$, $f_C$ is constructed by averaging more than 10 absorption images
		under the condition that almost all atoms localize on a sole sublattice.


\end{document}